\documentclass[twoside]{IEEEtran}
\usepackage{stfloats}
\usepackage{lipsum}
\ifCLASSINFOpdf
   \usepackage[pdftex]{graphicx}
   \usepackage{xcolor}
   \usepackage{soul}
   \usepackage{multirow}
\usepackage{comment}
\usepackage{amsmath,amssymb}
\usepackage{makecell}
\usepackage{caption}
\usepackage[colorinlistoftodos]{todonotes}
   
   \graphicspath{{../pdf/}{../jpeg/}}
  
   \DeclareGraphicsExtensions{.pdf,.jpeg,.png}
\else
\fi

\usepackage{amsmath}

\begin{document}
%
\title{IEEE 802.15.7-Compliant Ultra-low Latency Relaying VLC System for Safety-Critical ITS}
%
%
%

\author{Tassadaq~Nawaz, 
Marco~Seminara, %
Stefano~Caputo, %
Lorenzo~Mucchi,~\IEEEmembership{Senior~Member,~IEEE, }%
Francesco~Cataliotti, %
and Jacopo Catani

\thanks{
\copyright 2019 IEEE.  Personal use of this material is permitted.  Permission from IEEE must be obtained for all other uses, in any current or future media, including reprinting/republishing this material for advertising or promotional purposes, creating new collective works, for resale or redistribution to servers or lists, or reuse of any copyrighted component of this work in other works.}
\thanks{T.~Nawaz is with National Institute of Optics - CNR (CNR-INO), Sesto F.no, I-50019 Italy
e-mail: tassadaq.nawaz@ino.cnr.it .}
\thanks{M.~Seminara and F.~Cataliotti are with LENS and Physics and Astronomy Dept. - University of Firenze, Sesto F.no, I-50019 Italy
e-mail: seminara@lens.unifi.it, fsc@lens.unifi.it . }%
\thanks{S.~Caputo and L.~Mucchi are with the Dept. of Information Eng. (DINFO) - University of Firenze, Italy, e-mail: stefano.caputo@unifi.it, lorenzo.mucchi@unifi.it}
\thanks{J.~Catani is with National Institute of Optics - CNR (CNR-INO) and LENS, Sesto F.no, I-50019 Italy
e-mail: jacopo.catani@ino.cnr.it .}
}

%
%

\markboth{}
{Nawaz \textit{et al.}: IEEE 802.15.7-Compliant Ultra-low Latency...}
%



\maketitle


\begin{abstract}
The integration of Visible-Light Communications technology (VLC) in Intelligent Transportation Systems (ITS) is a very promising platform for a cost-effective implementation of revolutionary ITS and cooperative ITS protocols. In this paper, we propose an infrastructure-to-vehicle-to-vehicle (I2V2V) VLC system for ITS, implementing it through a regular LED traffic light serving as a transmitter and a digital Active Decode-and-Relay (ADR) stage for decoding and relaying the received information towards further incoming units. The proposed VLC system targets the challenging and important case of ultra-low latency ADR transmission of short packets, as this is needed for emerging applications of automatic braking, car platooning and other critical automatic and/or assisted driving applications. The experimental validation of the ADR VLC chain, as well as a thorough statistical analysis of errors distribution in the transmission, has been performed for short to medium distances, up to 50 meters. The performances of the designed system are evaluated by measuring the packet error rate (PER) and latency in the whole ADR transmission chain. Our analysis shows that our system attains ultra-low, sub-ms latencies at 99.9\% confidence level for PER as high as $5\times10^{-3}$, yet granting a latency below 10 ms even for distances of 50 m. The demonstrated system prototype is compatible with IEEE 802.15.7 standard.
\end{abstract}


%
\IEEEpeerreviewmaketitle

\section{Introduction}
In recent times, integration of intelligent communication devices in vehicles is significantly increased, aiming at reduction of fatality rates and injuries in urban scenarios. 
However, according to the World Health Organization report \cite{WHO1}, despite sizeable efforts in introducing active or assisted reaction capabilities to sudden events in last generation of vehicles, more than 1.2 million people annually died and 20 to 50 million injured in road accidents. This report further predicts that the traffic-related fatality rates would further increase and become the sixth largest cause of death in world by 2020 whereas it was the ninth largest in 1990 \cite{WHO1,WHO2}. Increasing the efficiency and safety of entire transportation system clearly poses the need for more advanced pervasive, low-latency vehicular interconnection  technologies, aimed at boosting the vehicle's active safety protocols capabilities in response to critical events. 

To cope up with this challenge, different types of vehicular communications, namely Infrastructure-to-Vehicle (I2V), Vehicle-to-Vehicle (V2V) and Vehicle-to-Infrastructure (V2I), are being explored. In particular, the emergence of IEEE 802.11p standard for short to medium range inter-vehicle communication, which amends the popular 802.11 protocol suite and the allocation of  dedicated frequency band for ITS in Europe  have provided the potential solution for future implementations of communication-based ITS safety applications \cite{VLCINITS3,VLCINITS4}. ITS connects vehicles, humans and roads through state-of-the-art information and communications technologies to increase the safety and efficiency of the transportation system and also to reduce the environmental pollution. The safety and efficiency of the road traffic can be substantially improved by enabling the wireless communications between I2V, V2V and V2I to share the information regarding their dynamical state (e.g position, speed, acceleration etc.) or information about real traffic situations (e.g traffic jams, accidents, critical events etc.) \cite{VLCINITS4,VLCINITS5,VLCINITS6}.



In this scenario, Visible Light Communications  (VLC), exploiting LED-based lighting systems  \cite{VLCINITS7}, recently raised as a very promising communication technology due to a combination of features lacking in common RF-based communication systems 
e.g. they are readily integrated in existing infrastractures and an optical information channel can be highly directional.
Exploiting such features could allow for the implementation of agile and highly-reconfigurable, ad-hoc subnetworks by realizing directional interconnectivity between users in both I2V and V2V configurations without the need for complicated packet structures. 

In general, VLC uses the visible light spectrum [400-790 THz] to provide for highly-integrable, energy-efficient wireless solutions targeting either high data rates or pervasive broadcast of short information packets with very low latencies, which is especially important in ITS safety-critical applications (see Sec. \ref{Motivation for latency}). This is achieved by exploiting the high modulation bandwidth possibilities given by new high-power vehicular and urban LED illumination sources to encode digital information in the optical carrier, with no alteration of human eye perception \cite{PHY1}.

Most of the initial work in the field of VLC is based on basic modulation techniques such as on-off keying and pulse position modulation \cite{PHY2}. IEEE 802.15.7  \cite{IEEE802.15.7} standard which was made public in 2011 has opted on-off keying and pulse position modulation as fundamental modulation schemes for outdoor applications. Moreover, Colour Shift Keying (CSK) is also considered as third modulation scheme. Focusing on VLC in ITS, two types of systems, namely, camera-based \cite{VLCSENSOR12,VLCSENSOR13,SURVEY85l,SURVEY86,SURVEY87} and photodiode-based \cite{SURVEY60,SURVEY64,SURVEY69,SURVEY70,SURVEY71,SURVEY72,SURVEY89} systems are considered in literature. The high speed cameras are expansive and computationally complex to be used in automotive industry.  On the other hand low cost photodetectors are quite efficient regarding noise performances and can be used for long distances.

In the perspective of introducing VLC as a robust and reliable technology in real ITS scenario, validating the low-latency and high-accuracy character of a IEEE 802.15.7-compliant VLC communication chain involving both I2V and V2V endpoints with real road signaling sources would be of high relevance for boosting the introduction of VLC in ITS applications, aimed, e.g., at minimizing road accidents and enable smart traffic management protocols in large cities.

 \begin{figure*}[!t]
 	\centering
 	\includegraphics[scale=0.35]{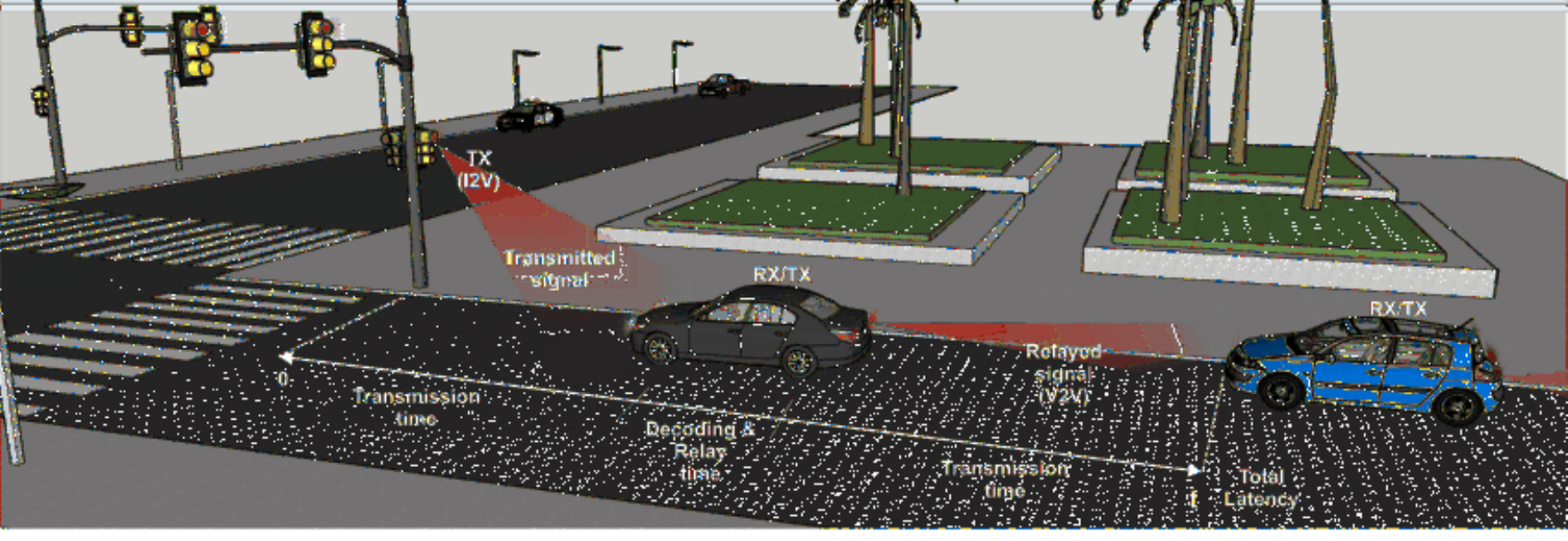}
 	\caption{Illustration of the proposed I2V2V ADR scenario: the traffic light sends a
 		message that is received by the first car (I2V) which acts as an active relaying node, decode the message and relay it to the car behind (V2V).}
 		\label{fig:relay_urban_chain}
 \end{figure*}

This paper presents an experimental implementation of a combined VLC Infrastructure-to-Vehicle-to-Vehicle (I2V2V) architecture for high-speed transmission of alert messages from a regular traffic light (TX) to an incoming unit (RX), which also embeds an ultra-fast active decode-and-relay (ADR) V2V stage towards further incoming units for ultrafast propagation of critical information through the vehicular chain (see Fig. \ref{fig:relay_urban_chain}). Our IEEE 802.15.7-compliant \cite{IEEE802.15.7} system is aimed at providing a cost effective solution for a short to medium range VLC to be employed in ITS outdoor safety-critical applications, and is based on a low-cost, open-source microcontroller platform (Arduino DUE) and attains for the first time ultra-low, sub-ms total latencies in the total TX-RX-ADR path through VLC. The experimental validation of our architecture, attaining baud rates as high as 230k with a Manchester encoding scheme, is performed using a regulated signaling infrastructure (a standard traffic light) for distances up to 50 m, featuring PERs approaching than $10^{-5}$. The PER performances of our system are statistically analyzed in order to retrieve average values for latency and packet delivery rate which can be used to reliably estimate automated reaction times to sudden events in ITS scenarios, paving the way towards a new generation of ITS and cooperative ITS implementations. In addition, a mathematical model of the latency is derived. The model is suitable to simulate the latency in real scenarios and design proper communication protocols and procedures for road safety in ITS applications. 

The paper is organized as follows. In Section \ref{Motivation for latency}, we discuss motivation for latency analysis. In Section \ref{VLC Transmission Chain}, we explain our system model including implementation details and performance metrics. Measurement campaign is discussed in \ref{sec:meas_campaign}. The experimental results and performances evaluation are shown in Section \ref{sec:experimental_results}. Finally, we draw some conclusions in Section \ref{sec:conc}.
 
 \section{Motivation for Latency Analysis for Safety Critical Applications}\label{Motivation for latency}
 The PER is a well established parameter used to evaluate  performances of a communication system. However, as pointed out in \cite{RENDA201626}, for active road safety application, PER alone could generally be insufficient to assess the awareness level of receiving units in case of sudden events, whilst a statistically-averaged latency value (SAL) (generalizing the Packet InterReception time (PIR) parameter in \cite{RENDA201626} to the case of ADR systems), taking into account the distribution of errors in the transmission chain as a function of relevant experimental parameters, could provide a much more reliable performance indicator. In ADR systems, the bare latency can be defined as the time interval elapsed between the first bit of transmitted message and the last bit of the relayed message after a correct reception and decoding process. To understand the importance of retrieving a SAL parameter, let us consider an example where a TX unit X transmits 100 packets to a unit Y in $1$ s, with a PER of $0.5$ corresponding to an arithmetically-averaged latency value of 20 ms. Let's now imagine two different error distribution scenarios: in the first case, in the transmission a good packet and a bad one are evenly alternating. In second scenario, the packets received at Y form clusters: $20$ packets are received in $0.2$ s, then there is no reception of the packets for next $0.5$ s, and the remaining $30$ packets are received in last $0.3$ s. From this example, it is evident that two scenarios differ enormously from road safety point of view, as in the former case Y has constant knowledge of X's status every 20 ms, in contrast to the second case where the Y's knowledge about X's status can be outdated for as much as $0.5$ s, with evident consequences for any automated response in case of sudden critical event. Neither PER nor a simple average of latencies keep into account clustering of errors, while the SAL is calculated from the observed statistical distribution of error clusters size and hence represents a much more reliable metric to determine a statistical response time and success rate for automated actions in intelligent vehicular networks. 
 \begin{figure}[!t]
 	\centering
 	\includegraphics[scale=0.28]{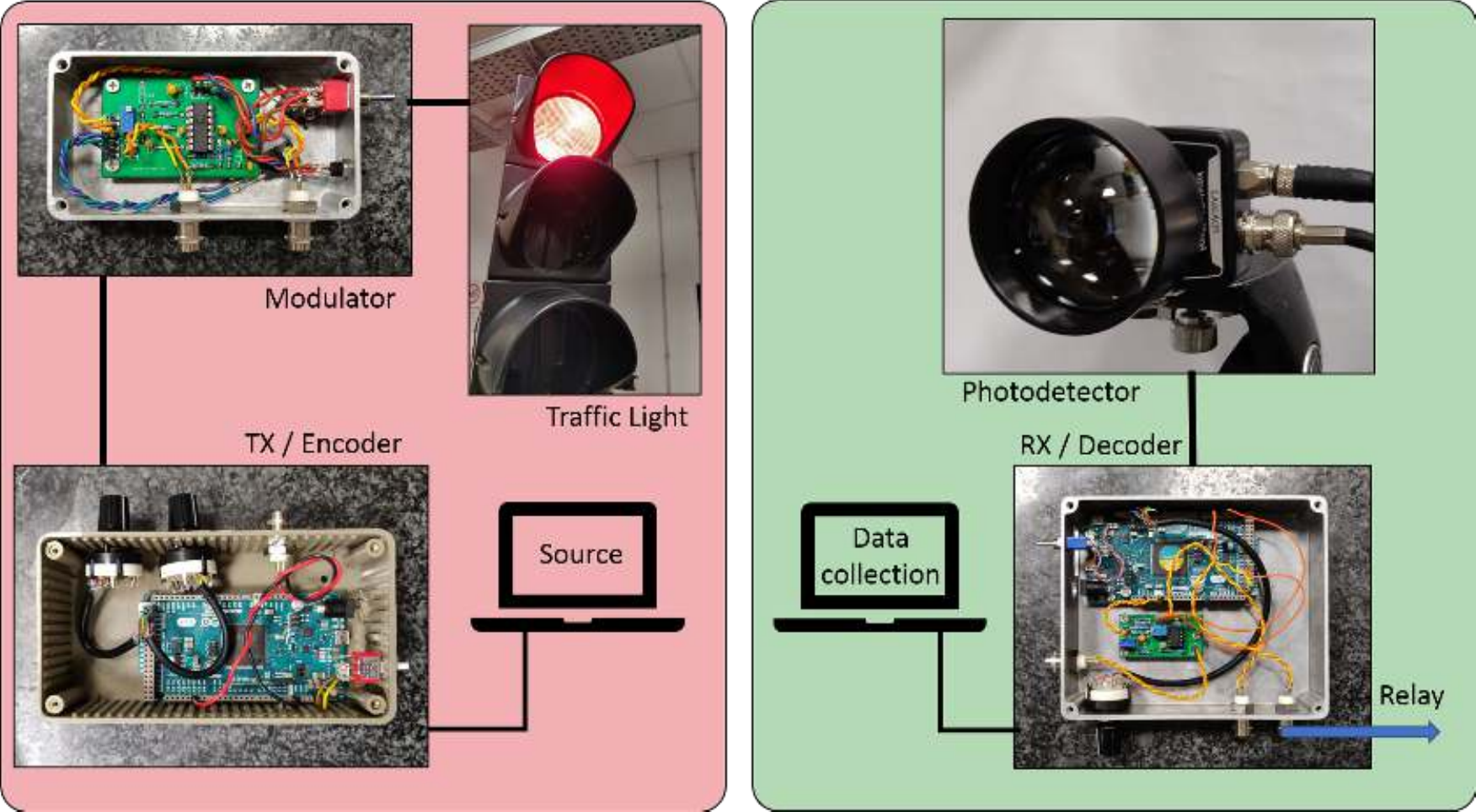}
 	\caption{VLC hardware blocks for our VLC active I2V2V decode and relay chain. Left (red): TX-modulator block, and regulatory traffic light source provided by ILES srl, Prato; right (green): RX-ADR block. Both blocks feature an Arduino DUE-based architechture as digital encoding and decoding engine.}
 		\label{fig_scheme}
 \end{figure}
  
 \section{VLC Transmission Chain} \label{VLC Transmission Chain}
 \subsection{Hardware overview}
 Our VLC I2V2V ADR system (Fig. \ref{fig_scheme}) is composed by a digital encoder/modualator, a regular traffic light, provided by ILES s.r.l, acting as TX light source, a fast, high-gain RX unit, a digitizer/decoder and an embedded ADR digital stage providing the possibility to modulate a further LED source (e.g. a rear lamp of a car) to relay received information to following units. A 1 Gs/s digital oscilloscope is used for signal recording and analysis. Details on the whole electrooptical TX-RX-ADR system will be given in a future work, as they are unnecessary to the scope of the present paper.
 \subsubsection{Transmitter}
 The TX stage is realized by adding current modulation to the LED lamp supply current. The modulation signal is provided by a digital encoder, realized through a microcontroller-based digital board (Arduino DUE). The TX stage transmits digital data up to 230 kBd by inserting digital information into the optical carrier emitted by the traffic light lamp. The On-Off Keying (OOK) is used for digital modulation, and Manchester encoding, obtained through a low-level, interrupt-based control of in-out ports of the board, is considered with OOK for data coding as it is recommended by IEEE 802.15.7 PHY 1 for outdoor VLC. The Manchester encoding guarantees a constant average signal allowing at the same time for a constant illumination by the traffic light, at the expenses of halving the effective bit rate of the communication chain. As the duty cycle of current driven into the LED source is 50\%, in order to maximize the modulation intensity whilst preserving the overall regulatory intensity of 100\%, we opted to perform a 0-200\%  modulation. This configuration did not lead to any derating in the LED source characteristcs in several months of continuous operation.
  \begin{figure}[!tb]
 	\centering
 	\includegraphics[width=0.9\columnwidth]{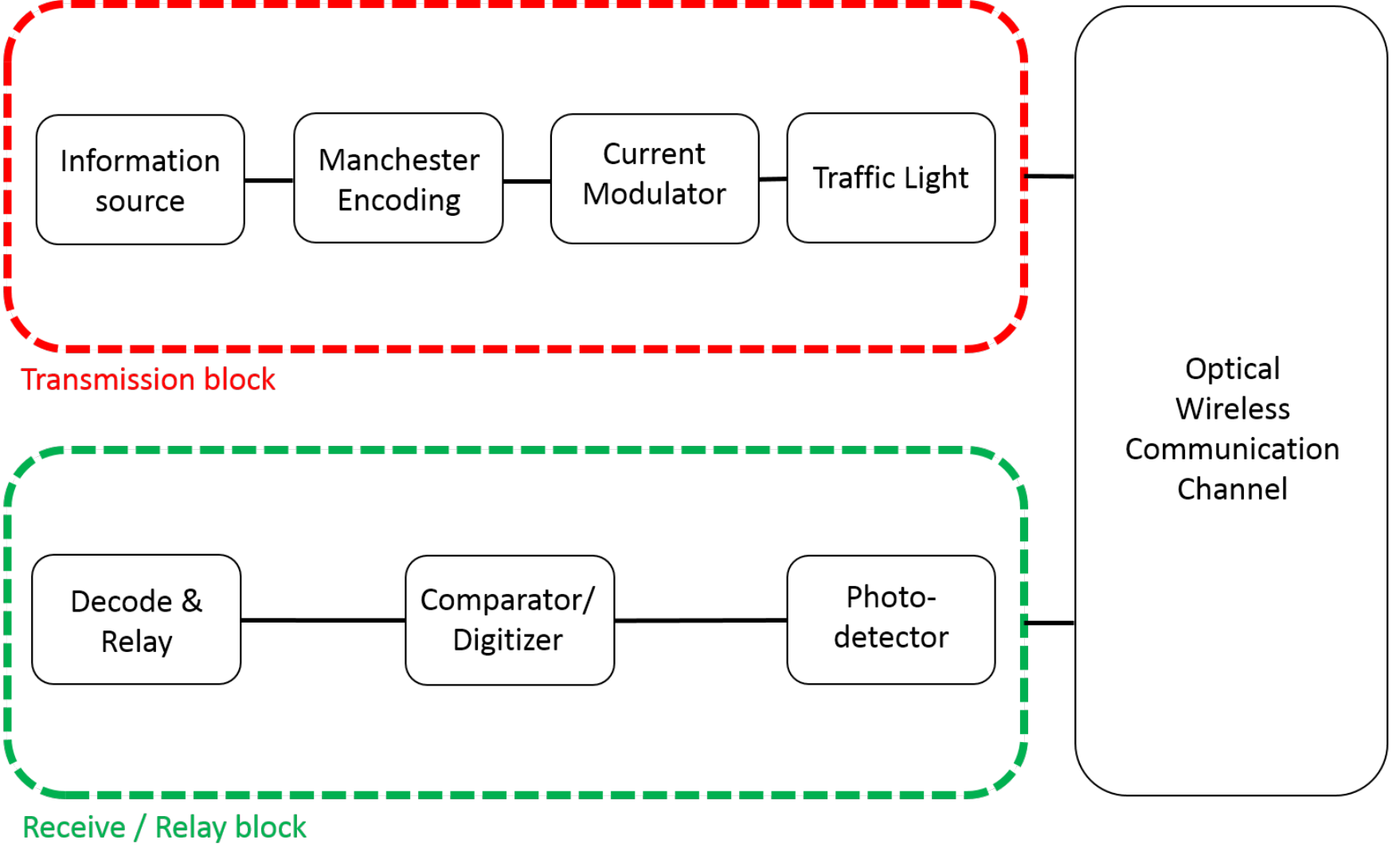}
 		\includegraphics[width=1\columnwidth]{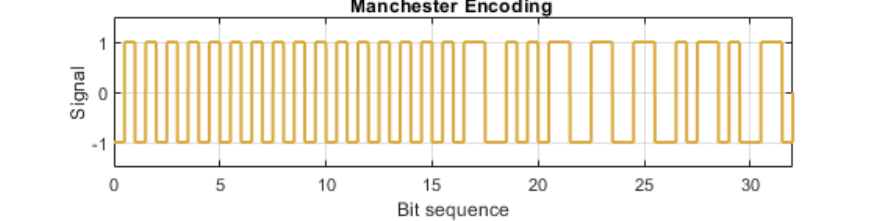}
 	\caption{Functional block diagram of our VLC Active Decode and Relay (ADR) chain. Lower panel reports the used Manchester bistream, composed by a 2-bytes pre-equalization and a 2-bytes data streams.}\label{fig:system_block}
 \end{figure}
  \subsubsection{Receiver/Decode and Relay}
 The RX unit collects the light from TX on a 36 mm$^2$ transimpedance photodiode, with variable secondary-stage gain. The collecting optics is an aspherical 2" uncoated lens, allowing for high optical gain, fundamental to have an efficient communication for high TX-RX distances. The photodiode is physically AC-coupled before the first transimpedance stage in order to reject spurious DC stray light components (such as sunlight or 100 Hz from artificial lights). As only the modulation component is retained, the gain value could be increased to high values (up to +50 dB for low baud rates) without risks of first-stage saturation. This amplification stage ensures a magnitude level of input signal higher than 20-40 mV even at 50 mt distances, which is considered as our limit for successful communication (see Sec. \ref{sec:experimental_results}).
 The amplified analog signal is then digitized by a
 variable-threshold comparator stage, and then analyzed and decoded by a digital RX board (based on Arduino DUE platform as well). Thanks to the relatively high computational power and speed of such platform, the decoded signal can be processed, re-encoded and delivered to a further modulation TX stage, eventually acting on  rear lamps/brake lamps for message relaying to following vehicles.

  \subsection{System model}
  \label{sec:system_model}
 
\begin{figure}[!tb]
	\centering
	\includegraphics[width=0.9\columnwidth]{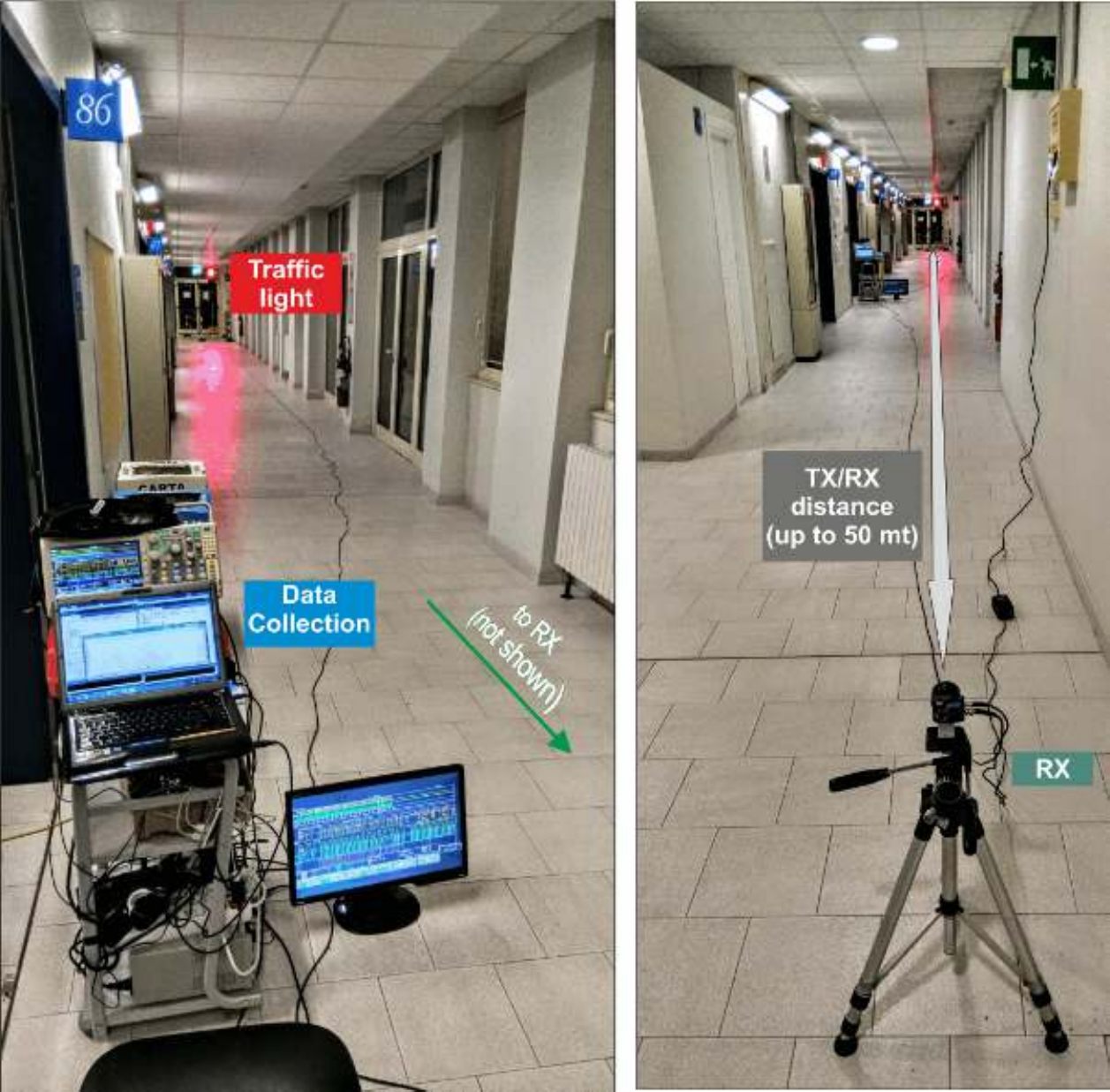}
	\caption{Experimental campaign: a standard traffic transmits VLC signal to a a photo-receiver stage equipped with digital Active Decode and Relay block. The data collection unit consists of a PC and an oscilloscope. The distance between transmitter and receiver is up to 50 m limited by available line of sight length in the building.}
	\label{fig:exp_campaign}
\end{figure}
 
The functional block diagram of the system is shown in Fig. \ref{fig:system_block}. The TX digital board produces the bit sequence (see lower panel of Figure) and packages it into a packet of 4 bytes (32 bits), 2 bytes preambles and 2 bytes of data (i.e. AL). Preamble follows a unique pattern (all 1's or 0's) that could be used for pre-synchronization between transmitter and receiver and also provide for useful signal equalization after transients. Then this data is fed into Manchester encoding block, providing the source signal for modulator, where data is inserted into the optical carrier using OOK current modulation, i.e. modulating the intensity of the traffic light, which represents the TX source.
The signal is passed through the optical transmission channel and received at photodetector, and then decoded and interpreted. The RX block presented in our prototype embeds an active relaying node: the Arduino DUE board decodes and bit-wise compares the received message with a stored reference message in less than 10 $\mu$s. If no errors are detected, the message is re-encoded and passed to a modulator for further clean relaying towards incoming units. We choose not to embed error correction algorithms as they are not inserted in the IEEE paradigms for outdoor VLC communications through 200 kHz carriers. Due to hardware limitations arising from interrupt conflicts, our ADR stage cannot receive bits while an active relaying of a previous data stream is occurring (see Fig. \ref{fig:broadcast}). We choose anyhow to keep the broadcast signal interpacket delay as short as possible (continuous broadcast) in order to limit signal level fluctuations due to long interpacket phase where no modulation is present. In this continous broadcast configuration, the number of relayed messages if no errors are detected cannot exceed 1/2 the number of transmitted messages even in the ideal transmission case, so we set PER = 1 when the number of relayed messages equals 0.5 the number of transmitted ones. In the beaconing case, where the interpacket delay is by construction much longer than the packet length, this correction factor is not needed (Fig. \ref{fig:beaconing}). In both cases, the  minimum attainable latency in the whole TX-RX-ADR process is highlighted in Figs.  (\ref{fig:broadcast}-\ref{fig:beaconing}), and attains 595 $\mu$s in both broadcast and beaconing configurations (see Sec. \ref{sec:latency_subsection}).

  \subsection{Performance Metrics}
  Prototype performance are evaluated according to the following metrics:
  
   \begin{figure}[!t]
 	\centering
 	\includegraphics[width=\columnwidth]{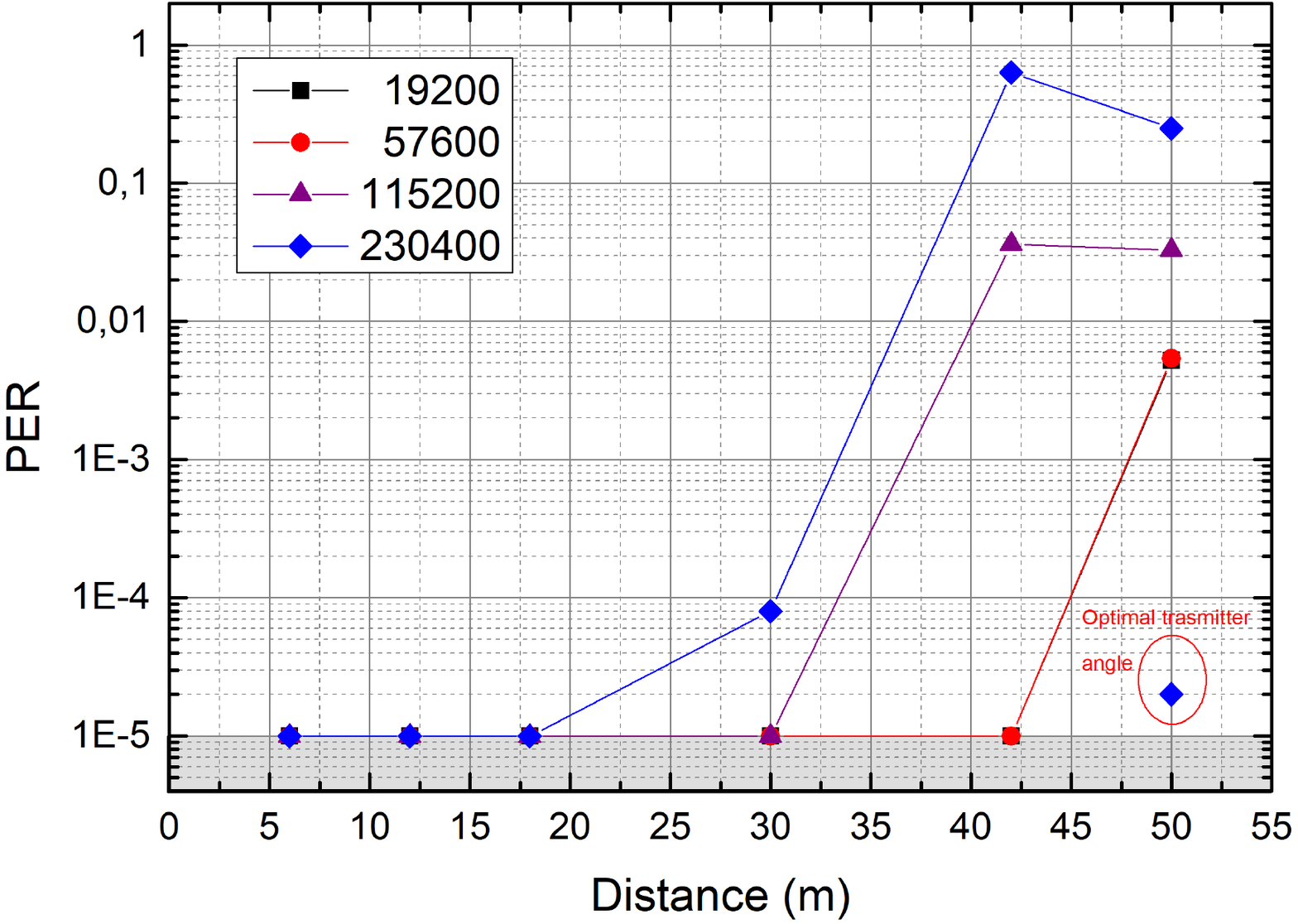}
 	\caption{PER performance of designed VLC I2V2V ADR system. PER is shown as a function of distance for various baud rates 230 kBd, 115 kBd, 57 kBd and 19 kBd. Shaded area in figure represents our upper limit in the observable PER ($10^{-5}$) due to transmitted number of packets. The encircled point at 50 m represents the absolute best performance obtained at our highest baud rate, if the traffic light vertical angle is optimized by few degrees.}
 	\label{fig:PER}
 \end{figure}
 
 \subsubsection{Packet Error Rate (PER)}
It is defined as ratio between lost VS total transmitted packets. A packet is considered to be lost if a single bit is altered during transmission. The lower observable PER ( $<10^{-5}$) is limited by the maximum number of transmitted packets ($2\times10^5$), chosen to limit the duration of the whole campaign to reasonable values. An estimation of BER is also possible in the hypothesis that a single bit is lost when an error is detected, which is very reasonable for low PER values. As our packets are composed with 32 bits, this leads to an upper value of BER of $3\times10^{-7}$.

\subsubsection{Latency}
With reference to Figs. (\ref{fig:broadcast}-\ref{fig:beaconing}), latency is defined as the time elapsed between the first bit of a TX message and the last bit of an actively-relayed message after the ADR block.

 \section{Measurement Campaign}
 \label{sec:meas_campaign}
 The measurement campaign is carried out in collaboration with ILES srl, a company producing and installing intelligent signaling elements, in the city of Prato - Italy. The measurements are performed in a long corridor (55 m) inside the Physics ad Astronomy Dept. of the University of Florence (see Fig. \ref{fig:exp_campaign}).  In experimental set up, the relative vertical height among photodetector and red lamp is set to implement a configuration where the collection lens is placed on the car dashboard ($\sim 105$ cm from ground \cite{2019arXiv190505019C}). 
 In this configuration, interference due to artificial and ambient light sources such as multipath reflections are always present, and require an equal or higher robustness in noise rejection schemes when compared to outdoor scenario implementations, where, e.g., the 100 Hz component is definitely less critical.
 
 The received signal is proportional to the optical flux collected by the condenser lens, hence decreases with increasing distance and incidence angle between optical axes of the transmitter and receiver optical elements. In this measurement campaign, the  photodetector optical axis is aligned towards the traffic light lamp center in order to maximize the amplitude of the received signal. In order to evaluate the performances of the designed prototype in various possible distance/gain configurations, the distance between traffic light and photodetector is changed from $6$ m to $50$ m with step sizes of $6$ to $12$ m, whilst the gain value is spanned in steps of 10 dB for all useful values, ranging from PER = 1 for too low or too high gain (limiting the RX electronic bandwidth to unsuitable values for a specific baud rate), up to the lowest detectable PER of $<10^{-5}$, limited by the number of transmitted packets (see above).

 \section{Experimental Results and Discussions}
\label{sec:experimental_results}
\begin{figure}[!t]
	\centering
	\includegraphics[width=\columnwidth]{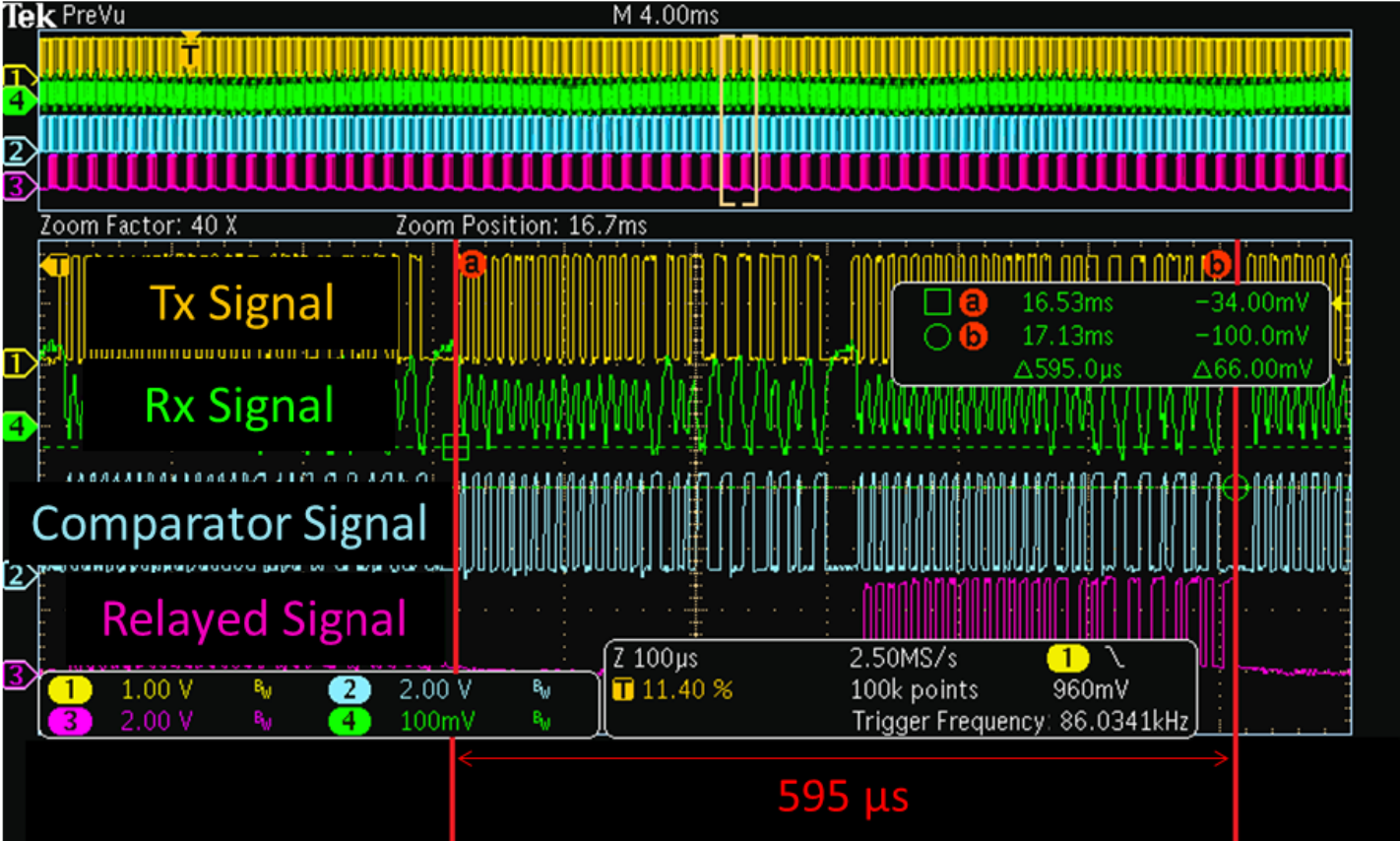}
	\caption{Continuous broadcast: oscilloscope screenshot (upper panel) with a zoom (lower panel) showing transmitted signal (yellow), received analog signal (green), digitized received signal (blue), and relayed signal after ADR block (purple). The red marker on the screenshot shows a minimum active packet relaying ($\bigtriangleup$) of 595 $\mu$s obtained  at a rate of 230kBd. Effects of artificial illuminations are visible in the analog green track as a 100 Hz residual oscillation. }
	\label{fig:broadcast}
\end{figure}
\subsection{PER Analysis}
For these experiments a predefined message is sent continuously at $230$, $115$, $57$, and $19$ kBd rates. Fig. \ref{fig:PER} shows the dependence of the PER performance on the distance between traffic light and photodetector for various baud rates. The PER performance of the system is decreased as the distance between the transmitter and receiver is increased due to reduction of the received modulation signal with respect to the residual stray ambient components and line noise, which are relatively independent on the RX-TX distance. Our measurements demonstrate that our VLC-based prototype is able to establish communications up to the maximum available distance of $50$ m for all of the tested baud rates. A  PER value $<10^{-5}$, corresponding to a lossless transmission in our observation window, is obtained for distances up to $42$ m for $19$ and $57$ kBd. Higher baud rates of $230$ and $115$ kBd achieve lossless  transmission up to $18$ and $30$ m, respectively. The ultralow-latency configuration (230 kBd) features PER $<10^{-4}$ at $30$ m and still grants PER$<0.3$ at 50 m. Noticeably, anyhow, near-lossless PER value of $2\times10^{-5}$ is recovered at 50 m even for 230 kBd if the traffic light is slightly tilted backwards by few degrees. Indeed, as verified in a previous work of ours \cite{2019arXiv190505019C}, roadside traffic lights are designed to maximize visibility (and so the VLC signal) in the range of 12-18 m. In case long distances should be privileged for casting long-range VLC signals, hence, it is beneficial either to recline the lamp optics of roadside traffic lights by few degrees in order to redirect the shaped beam towards higher distances, or to prefer the usage of over-road traffic-lights.  

\begin{figure}[]
	\centering
	\includegraphics[width=\columnwidth]{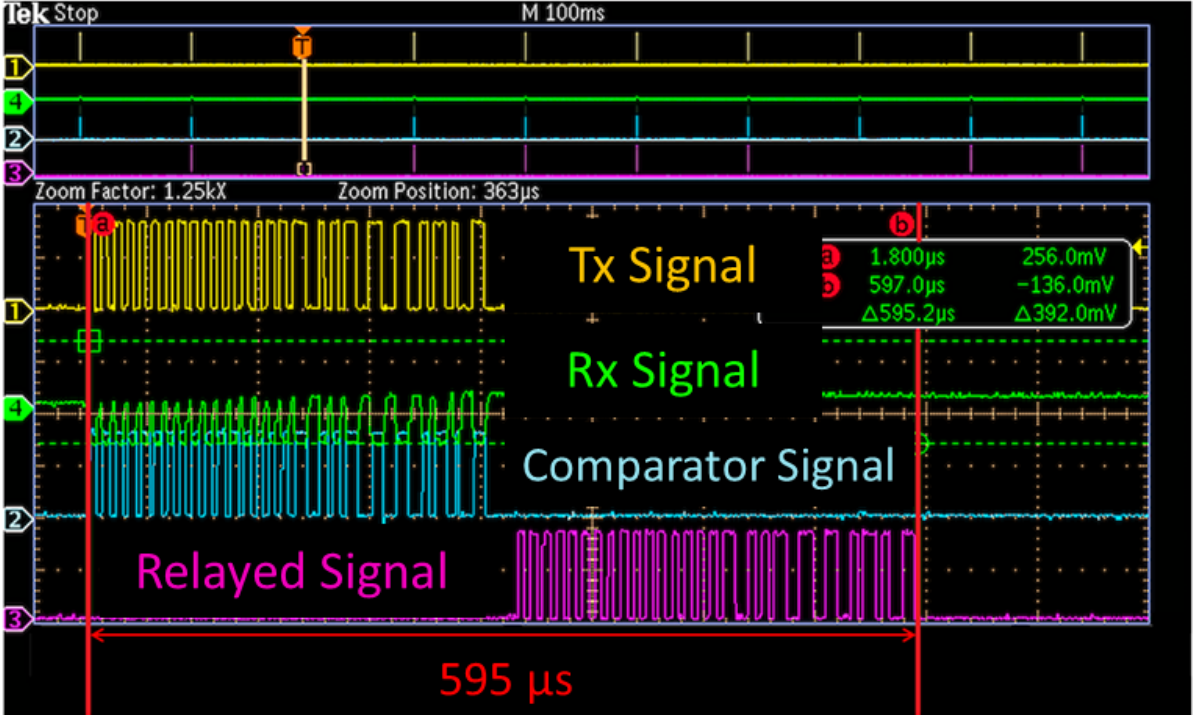}
	\caption{Beaconing configuration: oscilloscope screenshot (upper panel) with a zoom (lower panel) showing transmitted signal (yellow), received analog signal (green), digitized received signal (blue), and relayed signal after ADR block (purple). The red marker on the screenshot shows a minimum active packet relaying ($\bigtriangleup$) of 595 $\mu$s obtained  at a rate of 230kBd. }
	\label{fig:beaconing}
\end{figure}

Without optimal lamp orientation, the PER performances at 50 m remain good ($ 0.007$) for both $19$ and $57$ kBd. In practice, a drop into the RX-ADR block capabilities to correctly receive packets is found around 20-40 mV in the signal level. This value is mainly arising from 100 Hz interfering signal at the receiver, resulting as a residual neon light component after the first AC filtering stage. We found not beneficial to further increase the AC decoupling frequency, as this would start to cut the modulation signal as well. Fig. \ref{fig:broadcast} is an oscilloscope screenshot of the data stream across various blocks, taken in real time during experiments. In this figure yellow colour shows the transmitted signal, the green colour presents the received analog signal after the photodetector, blue is for digitized received signal after comparator and the purple track shows relayed packets. The impact of the artificial lights on received signal shape can be observed as the residual modulation appearing in the green track of upper panel in Fig. \ref{fig:broadcast}. Incidentally, this could be the origin of the deviation of PER vs distance from monothonic behavior at very low signals in Fig. \ref{fig:PER} (i.e. distances $> 40$ m and high baud rates). Indeed, as the light coming from lamps on the ceiling can enter more or less the detector field of view depending on its position along the corridor, the consequences of this stray 100 Hz signal can start to have an effect on PER at very large distances, where the signal is much feebler. 

We remark that the better experimental values obtained of PER $<10^{-5}$ and estimated BER below $3\times10^{-7}$ represent a worst-case limit due to limitations of number of transmitted packets. From Fig. \ref{fig:PER} we have clear indications that, especially at distances up to 30 m, actual PER and BER limit values are much lower than our upper limits.

\subsection{Ultra-low ADR latency performances in broadcast and beaconing configurations}\label{sec:latency_subsection} 
Besides the good PER and BER performances shown for distances up to 50 m, another remarkable feature of our VLC prototype is the demonstration of ultra-low, sub millisecond active decode and relaying latencies (see red markers inf Figs. (\ref{fig:broadcast}-\ref{fig:beaconing})): a data packet is transmitted, decoded, compared and relayed in 595 $\mu$s for IEEE-compliant modulation frequencies of $\sim 200$ kHz (230 kBd in our case). This achievement makes our low-cost, highly-integrable VLC architecture suitable for time-critical, road-safety ITS applications,  Further, such low latency values indicate that this architechture would fulfill the latency standards of 5G technology in IoT applications.
\begin{figure}[!tb]
	\centering
	\includegraphics[width=1\columnwidth]{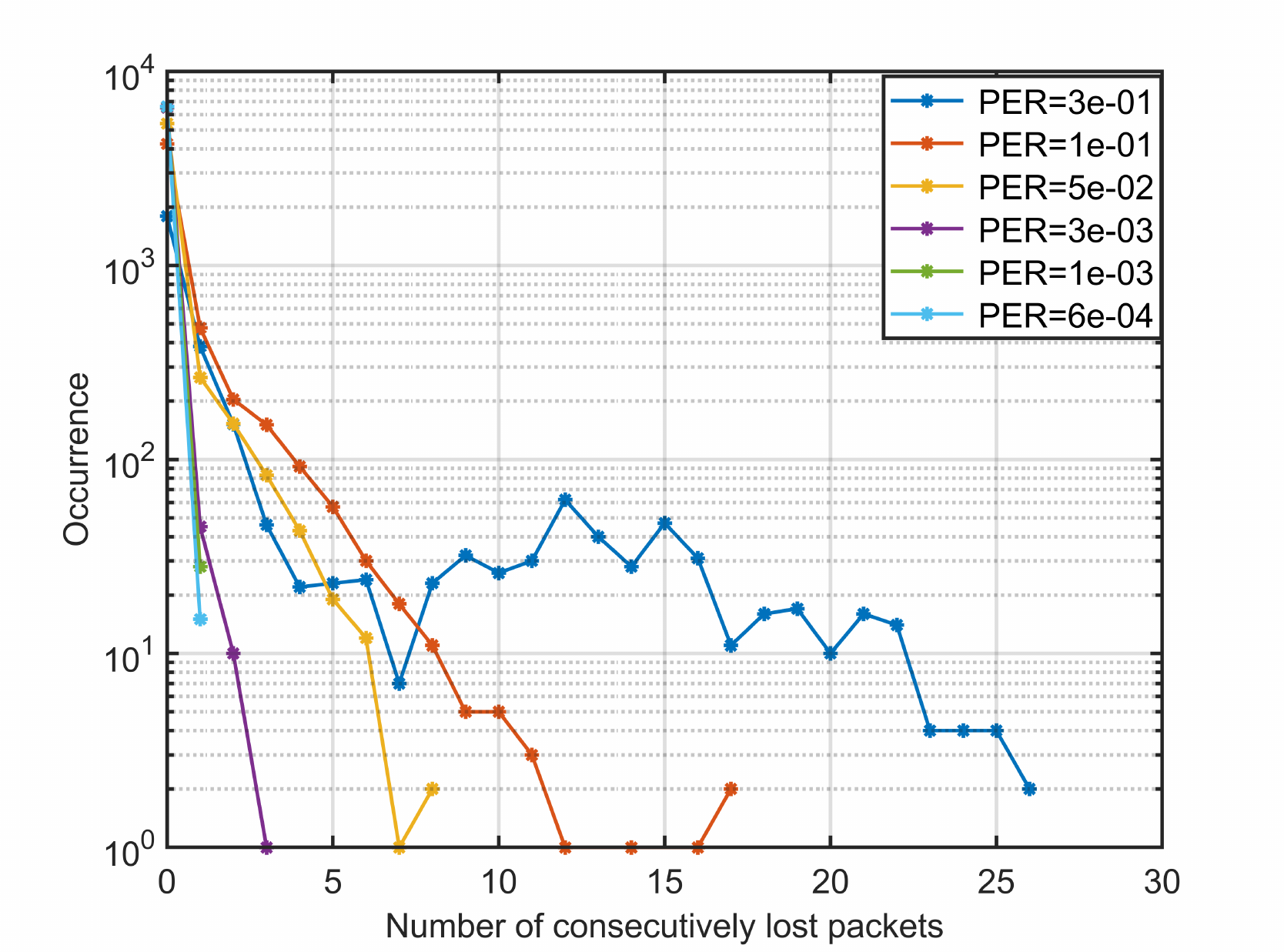}
	\caption{Occurrence vs number of consecutively lost packets (i.e. the cluster size), retrieved from data obtained during experiments for various PERs. Legend presents different PERs considered for latency analysis.}
	\label{fig:occurrence}
\end{figure}
As a proof of principle, we also test the suitability of our system for beaconing of situational information in one example configuration (230 kBd). The beaconing interval is set to be much longer (100 ms) than the packet time ($\sim 280\,\mu$s), and this constitutes a major difference with respect to the broadcast case, as the average amplitude of detected datastream is now much less constant due to signal transients driven by long interpacket delays. The beaconing performances are shown in the Fig. \ref{fig:beaconing}, and a sub-millisecond (595 $\mu$s) relaying is also achieved in this case, without appreciable loss of packets. Figs. (\ref{fig:broadcast}-\ref{fig:beaconing}) highlight that the prototype presented in this paper is suitable for both event-triggered message broadcast and continuous data transmission.

\subsection{Statistical Latency Analysis}\label{sec:latency_statitics}

As discussed in Sec. \ref{Motivation for latency}, neither bare minimum latency values corresponding to a certain baud rate, nor the PER alone can provide an effective metric to determine a most probable successful data delivery time in ITS applications. Since (especially for high PER values) clusters of consecutively lost packets can appear, a  statistically-averaged latency (SAL) should be rather retrieved from a statistical analysis of the lost packets distribution. This section presents an analysis of such distributions, recorded for different PER values. Differently from PER measurements, in such measurements set we had to acquire whole tracks in order to perform an accurate post-processing analysis of errors distributions and latencies. The memory depth of the oscilloscope sets the acquisition length to a maximum amount of 10000 sent packets, and since we have to perform an analisys on statistically-relevant datasets, we restrict the analysis to PER values not lower than $\sim10^{-3}$, in such a way that the total recorded number of lost packets in our observation window exceeds 10, considered as a limit value for a statistically-relevant error dataset. For lower PERs, indeed, errors distributions can be strongly affected by the finite-size nature of our ensemble, failing to provide for inferential capability to our statistical analysis. 

\begin{figure*}[!t]
	\centering
	\includegraphics[width=2\columnwidth]{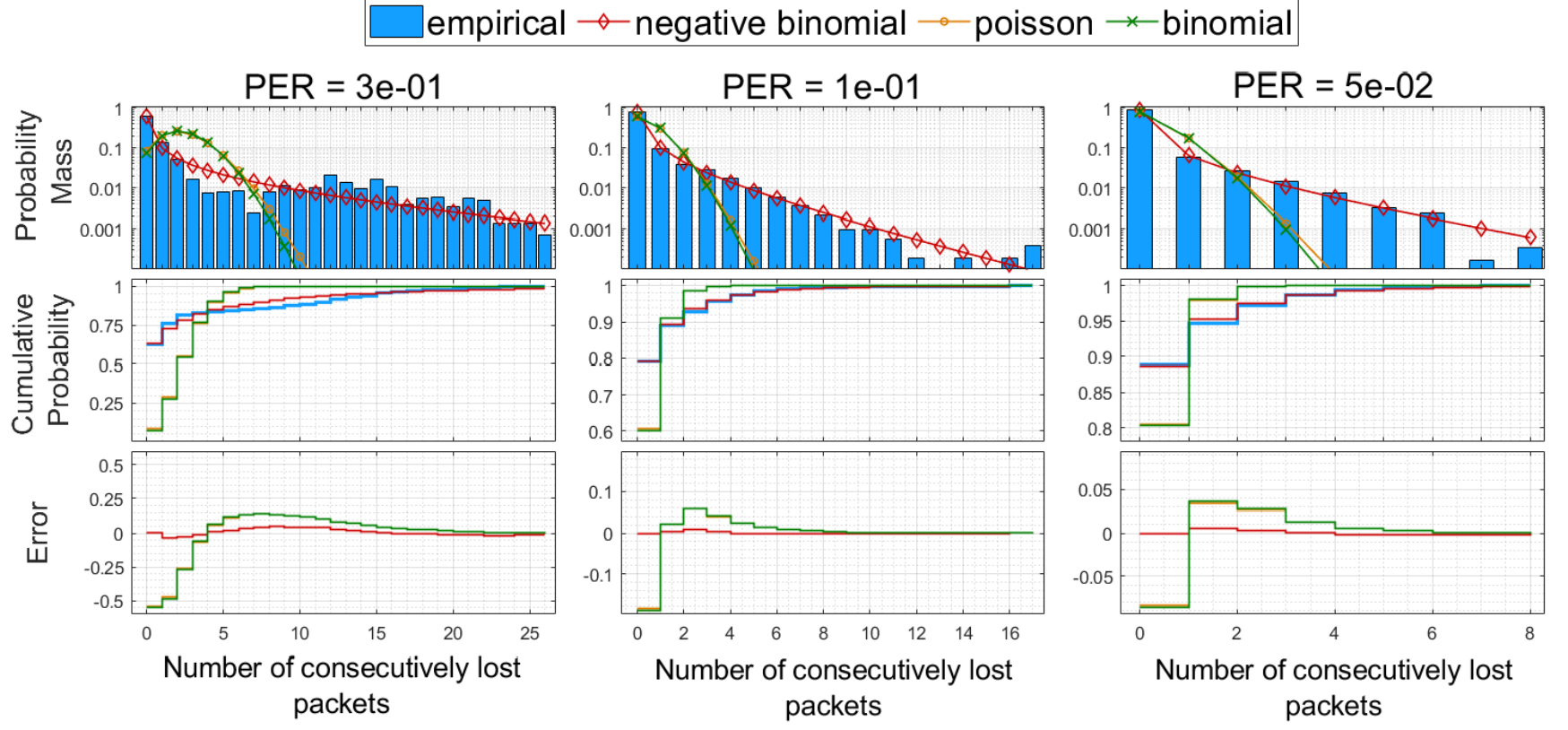}
	\caption{Latency PMF and CDF with curve fitting for different lost-packets distribution models (Negative binomial, Poisson, and binomial), along with fitting error (lowermost panel) for significant PERs (see text for details).}
	\label{fig:distributions}
\end{figure*}

The curves for number of occurrences vs. number of consecutively lost packets (cluster size) for different PERs are plotted in Fig. \ref{fig:occurrence}. It shows that the clustering in relaying messages is increased for higher values of PER. The maximum observed cluster size is $26$ in case $3\times10^{-1}$, whereas it is $17$ for  $1\times10^{-1}$, $8$ for $5\times10^{-2}$ and $3$ for $3\times10^{-3}$. For three PER cases (see Fig. \ref{fig:distributions}) we extracted the corresponding probability mass function (PMF) and cumulative distribution function (CDF) \cite{pap}, reporting the results in Fig. \ref{fig:distributions}. The clustering of errors tends to increase as the PER is increased, still featuring a smooth trend as a function of the cluster size. In order to infer a best estimation for SAL, we tried to identify a most suitable model for clusters distribution testing three different fit models (binomial, negative binomial, and Poisson) against our data. 

Fig.~\ref{fig:distributions} shows the PMFs and the CDFs of the empirical data as well as three discrete distributions: binomial, negative binomial and Poisson. The same figure also reports the CDF error, i.e., the difference between the CDF of empirical data and the CDF of the above mentioned distributions. 

As Fig. \ref{fig:distributions} shows, we obtain a very good agreement with negative binomial hypothesis for all of the analyzed PER values. 
The PMF of a negative binomial (or Pascal) distribution is 
\begin{equation}\label{eq.pascal}
    f(k,r,p) = \binom{r+k-1}{p} p^k (1-p)^r
\end{equation}
where $r$ is the number of failures, $k$ is the number of successes, and $p$ is the probability of success. 
This agreement enables us to use the negative binomial hypothesis in order to make predictions on system level performances of our implementation from the SAL point of view as a function of PER (Sec. \ref{sec:SAl_estimation}).

\begin{figure}[!tb]
	\centering
	\includegraphics[width=1\columnwidth]{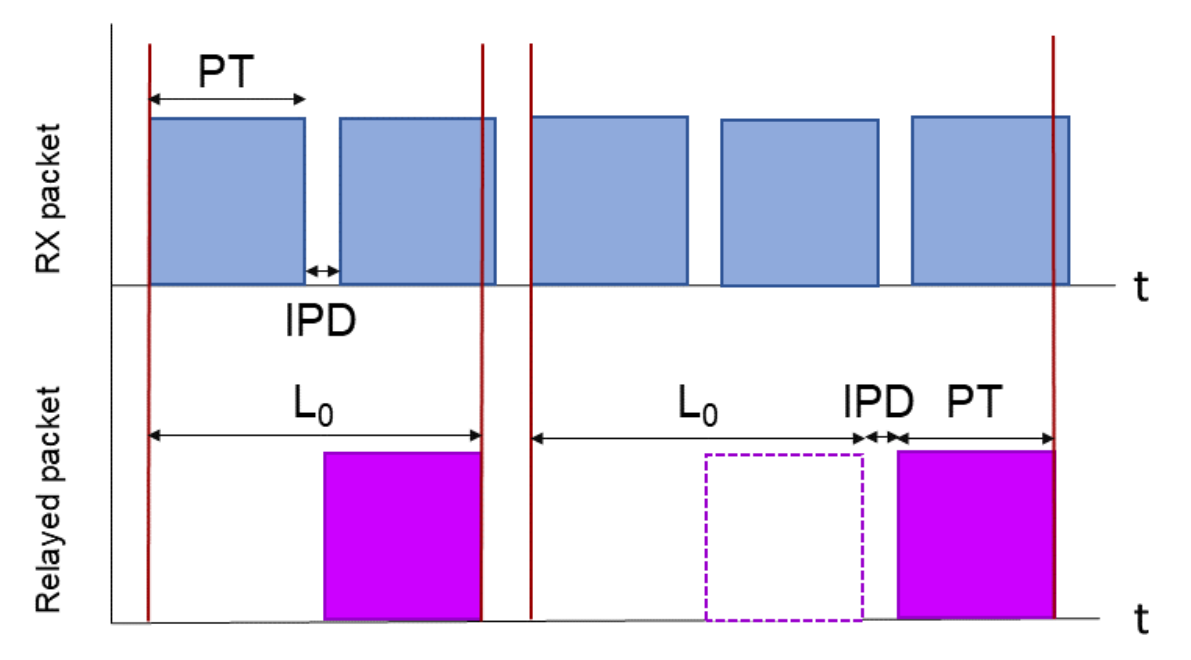}
	\caption{Latency reconstruction from packet structure:  $L_0$ is the minimum time required to relay a packet, $N_{Lost}$ is the number of consecutively lost packets, $IPD$ is interpacket delay, and $PT$ is the packet size in time.}
	\label{fig:latencycalculations}
\end{figure}

In order to obtain a SAL it is first of all important to connect the CDF of error cluster size (central panels of Fig. \ref{fig:distributions}) to its effect in terms of communication delay. 
With reference to Fig.~\ref{fig:latencycalculations}, the latency can be inferred as a function of error cluster size (retrieved from raw data) using following equation:
\begin{equation}
L_n=L_0+N_{Lost}(IPD+PT)
\label{eq:latency}
\end{equation}
where $L_n$ is the total latency, $L_0$ is the minimum time required to relay a packet, $N_{Lost}$ is the number of consecutively lost packets (cluster size), $IPD$ is interpacket delay, and $PT$ is the packet duration. Parameters $L_n$, $L_0$, $IPD$ and $PT$ appearing in Eq. \ref{eq:latency} can be considered as constants, with observed deviations on the timescale of $10\,\mu$s.
\begin{figure*}[!t]
	\centering
	\includegraphics[width=2.2\columnwidth]{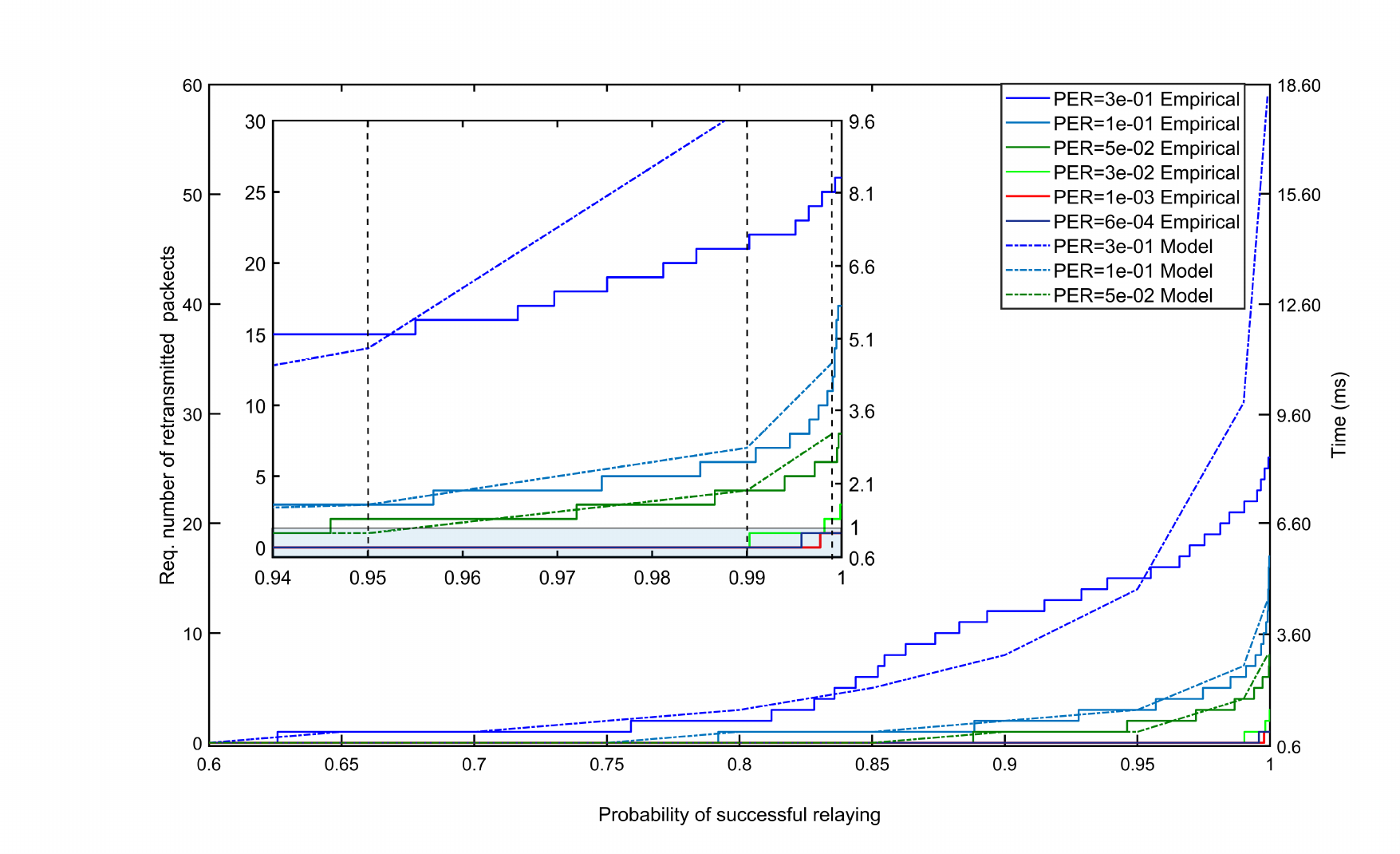}

	\caption{Performance of designed prototype for road safety critical applications. Right axis is calculated for 230 kBd, see Eq. \ref{eq:latency}. Inset image is a zoom in of the probability of successful relaying from 0.94 to 1. The vertical dash lines are used to represent the PSR of 0.95, 0.99 and 0.999. Dash-dot lines report the prediction of negative binomial model (see text).}
	\label{fig:success_prob}
\end{figure*} 

Importantly enough, the cluster size CFD analysis, performed for several PER configurations, delivers the required number of broadcast packet retransmissions granting a certain success probability in our TX-RX-ADR transmission chain (see solid lines of Fig. \ref{fig:success_prob}). In turn, through Eq. \ref{eq:latency}, we can link this packet number to the corresponding latency (right axis of Fig.  \ref{fig:success_prob}, obtained inserting time parameters for 230 kBd).

The dashed vertical lines in the inset indicate three significant success probability levels (95\%, 99\% and 99.9\%), whilst the horizontal line highlights the sub-ms latency regime. Most significantly, even in the worst PER case ($3\times10^{-1}$), at 230 kBd, a successful transmission can be accomplished with 99.9\% confidence level in only $\sim8$ ms, whilst the sub-ms relaying regime (highlighted by the shaded area in the inset of Fig.~\ref{fig:success_prob}) is granted already for PER values $\lesssim 10^{-3}$  (red line). When combined to results reported in Fig. \ref{fig:PER}, these result noticeably demonstrates, with a confidence level better than 0.999, that: a) our VLC system is able to perform successful communication up to 50 m at 230 kBd and perform ADR of information in less than 10 ms; b) successful communication with ultra-fast, sub-ms ADR performances is granted up to more than 30 m, where the measured PER @ 230 kBd attests well below $10^{-3}$. 

\begin{figure}[h]
    \centering
    \includegraphics[width=0.99\columnwidth]{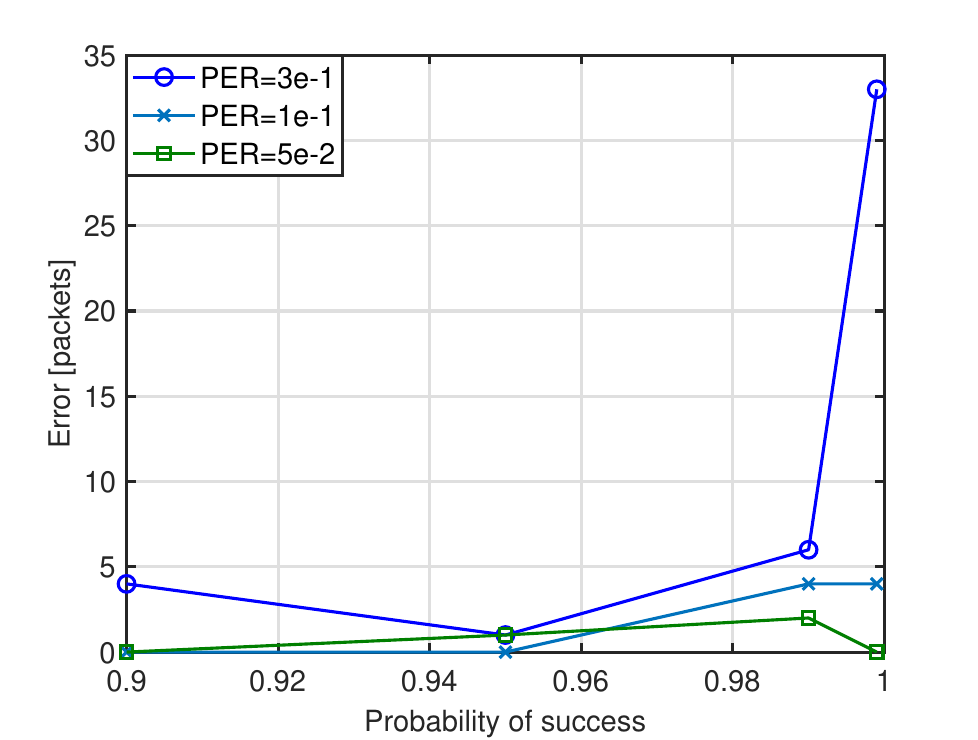}
    \caption{Difference between the predicted number of required packets and the measured values for three relevant PER values as a function of transmission success probability (see Fig. \ref{fig:success_prob}).}
    \label{fig:error}
\end{figure}

The best fitting model (negative binomial) for $N_{Lost}$ (dash-dot lines) can analogously be used to infer $L_n$ as a function of cluster size and to extract the SAL value from knowledge of cluster size distribution for different PER values.
As it can be seen, the latency values of the proposed model and the empirical data are close, and the discrepancy is always $\leq 5$\,packets for probability $<$ 0.95 (see inset of Fig. \ref{fig:success_prob} and Fig. \ref{fig:error}). 
Only in case of higher probabilities and high PER ($3\times10^{-1}$) the accuracy appears to get lower, as the predicted cluster size value is significantly higher than the measured value. Notwithstanding this model discrepancy, for which we did not find a trivial explanation addressing it to the finite-size nature of our experimental sample, the model proves to provide correct predictions for error distributions in a very wide set of transmission parameters.


\begin{figure}[h]
    \centering
   \includegraphics[width=0.99\columnwidth]{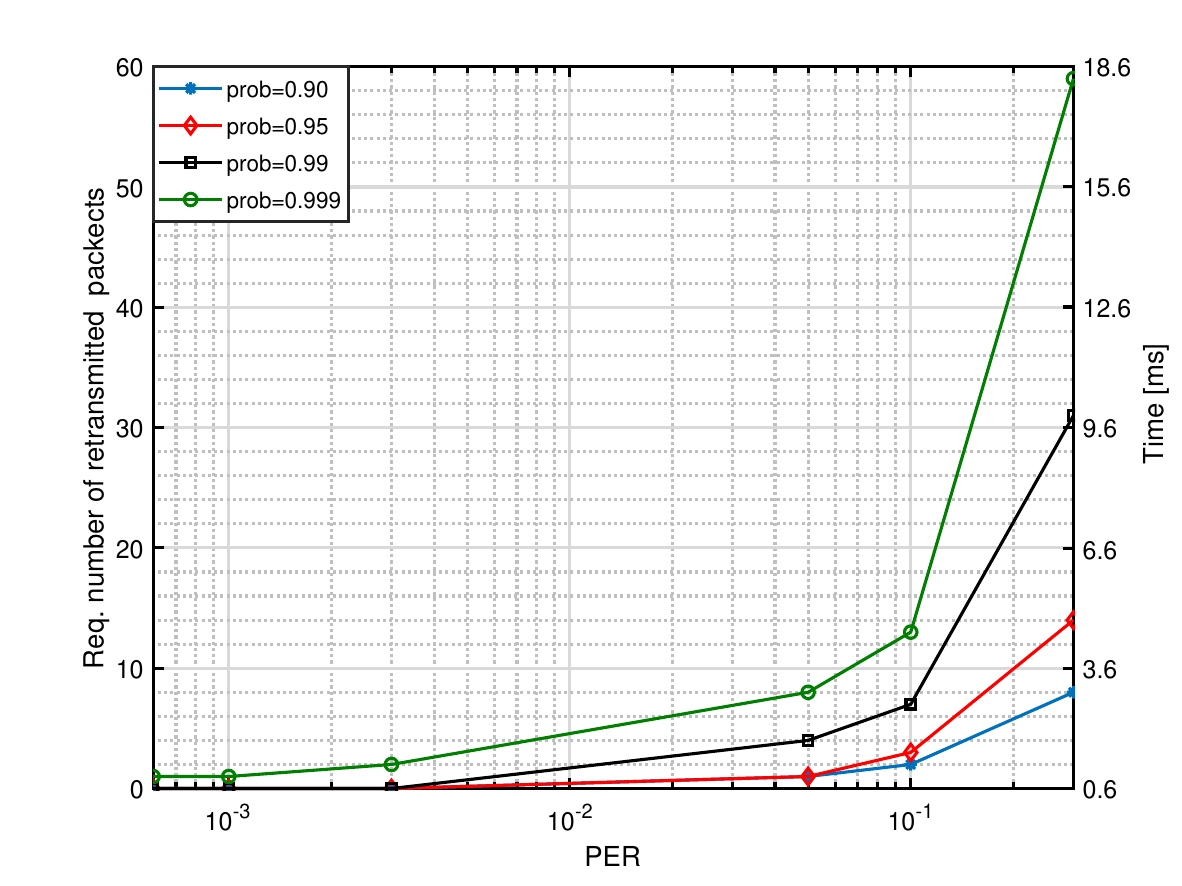}
    \caption{Model predictions for latency (SAL) as a function of the PER for different probability of successful relaying.}
    \label{fig:PERvsLATENCY}
\end{figure}

\begin{table}[h]
\label{tab:latency}
  \renewcommand*{\arraystretch}{1.4}
    \centering \caption{Parameters of best-fitting PMF model for latency.}
    \begin{tabular}{|c|c|c|c|c|c|} \hline 
        PER & Parameters & \multicolumn{4}{|c|}{Target probability}\\ \hline
         & & 0.9 & 0.95 & 0.99 & 0.999 \\ \hline 
         & & \multicolumn{4}{|c|}{Latency [no. of packets]} \\ \hline
        $3\times 10^{-1}$ & r=0.1691; p=0.0638 & 8 & 14 & 31 & 59\\ \hline
        $1\times 10^{-1}$ & r=0.1719; p=0.2555 & 2 & 3 & 7 & 13 \\ \hline 
        $5\times 10^{-2}$ & r=0.1089; p=0.3342 & 1 & 1 & 4 & 8\\ \hline
        $3\times 10^{-3}$ & r=0.028; p=0.7053 & 0 & 0 & 0 & 2\\ \hline
        $1\times 10^{-3}$ & $\lambda$=0.0042 & 0 & 0 & 0 & 1\\ \hline
        $6\times 10^{-4}$ & $\lambda$=0.0023 & 0 & 0 & 0 & 1\\ \hline
    \end{tabular}\label{tab:latency}
\end{table}
The latency model parameters corresponding to dash-dot lines of Fig.\,\ref{fig:success_prob} are also reported in Table \ref{tab:latency}. It is worth to note that for the cases of very low PER (few errors in the measurements) the best fitting PMF is rather given by a Poisson distribution with $\lambda$ parameter reported in Table~\ref{tab:latency}. Similarly to the higher PER and higher success probability case, the discrepancy can exceed 5 packets. The error of the model is depicted in Fig.~\ref{fig:error}. 

\subsection{Model estimation of SAL VS PER}\label{sec:SAl_estimation}

By using the best fitting model, we can derive the expected latency (SAL) corresponding to a successful decode-and-relay transmission process with a target probability, in generic communication conditions corresponding to a certain PER. Fig.~\ref{fig:PERvsLATENCY} shows the estimated SAL expressed as number of packets as a function of PER for different target probability of successful relaying. Right vertical axis shows the mapping on latency time obtained for 230 kBd by using Eq. \ref{eq:latency}. The latency grows monotonically with PER and target probability, but remarkably, even in the case of relatively high PER values ($10^{-1}$), which are far from suitable for internet links, the model predicts for our sytem ADR latencies below 10 ms even for target success probabilities of 99.9\%, whilst the sub-ms regime is expected already for PER $\lesssim 5\times10^{-3}$.

\begin{table*}[h]
\caption{Reaction and stop distance for an average family car with good tires over dry asphalt. \\ VLC Vs IEEE802.11p Vs human latency.}
  \renewcommand*{\arraystretch}{1.4}
\centering 
\begin{tabular}{|c|c|c|c|c|c|c|c|c|c|c|c|} \hline
\thead{v\\ \,[km/h]} & \thead{D\\ \,[m]} & \thead{PER\\@57\,kBd}  & \thead{Reaction \\Latency\\ @99\% [ms]} & \thead{Relay \\Latency\\ @99\% [ms]} & \thead{Brake \\distance [m]\\ ($\mu = 0.7$)} & \multicolumn{3}{c|}{Reaction dist. [m]} & \multicolumn{3}{c|}{Stop dist. [m]}  \\ \hline
      &          &       &         &  &  & VLC  & IEEE802.11p  & Human  & VLC & IEEE802.11p & Human \\ \hline
40    & 10        & $\sim 10^{-5}$ & $\sim 1.2$  & $\sim 2.4$  & 9.00  &  0.01    &   1.11           &  15.22      & 9.01    &   10.11          &  24.22     \\ \hline
60    & 20       & $\sim 10^{-5}$ & $\sim 1.2$  & $\sim 2.4$ & 20.25   &   0.02   &    1.67          &   20.83     &  20.27   &   21.92         &   41.08    \\ \hline  90  & 45       & $\sim 10^{-4}$ & $\sim 1.2$  & $\sim 2.4$ & 45.55  &  0.03    &    2.50          &   34.25     &   45.58  &    48.05         &     79.80 \\ \hline
\end{tabular}\label{tab:veh}
\end{table*}

\subsection{Application to road safety}
It is interesting to give a realistic estimation of the possible advantages in terms of road safety which could be obtained by introducing our low-latency architecture in realistic road scenarios. To this scope we give an estimation of the reduction in the stopping distance of standard vehicles in case of critical events (traffic light suddenly turns red) as compared to the human reaction case. 

The total stopping distance is the sum of the perception-reaction distance $D_{p-r}$ and the braking distance $D_{brake}$ \cite{wong1993theory}
\begin{equation}
    D_{stop} = D_{p-r}+D_{brake}=v t_{p-r}+\frac{v^2}{2 \mu g}
\end{equation}
where $v$ is the vehicle speed, $t_{p-r}$ is the reaction time, $\mu$ is the friction coefficient and $g$ is the gravitational acceleration. A road in good condition usually shows a friction coefficient of $0.7$.  

As a general remark, since at large distances the PER value for the lowest examined baud rates (19 and 57 kBd) is orders of magnitudes better than for 230 kBd, we assume that in long-cast safety-critical message delivery applications ($\gtrsim$ 30 m) it could be more effective to employ lower, more reliable baud rates, as such low PER values would correspond to a much lower SAL value even if the PT and IPD times are 4 times larger. Since long-distance performances of our system are equivalent at 57 and 19 kBd (see Fig. \ref{fig:PER}), in the following analysis we will use 57 kBd as transmission rate.

Table~\ref{tab:veh} shows a comparison between reaction and total stopping distances of an average family car in a dry road calculated for three relevant initial speed and distance-to-traffic-light situations. Different distances correspond to different PERs, and hence different latencies (see previous section). The reaction/stopping distances are evaluated by considering the reaction time equal to latency (delay) in the human-, IEEE802.p- and VLC-triggered braking case. Tests in real conditions found the fastest reaction time of car drivers to be $t_{p-r}=1.37$\,sec \cite{reaction}, whilst the latency requirements in outdoor road applications are given in \cite{5514475,8347079}. In particular, IEEE802.11p standard provides a maximum latency of $100$\,ms for RF-based road safety applications. For the VLC case, the reaction Latency, i.e. the time interval occurring between the beginning of a transmitted packet and the first correct packet reception at the receiving unit (without relaying) can be obtained by simply subtracting to ADR latency the Packet Time (see Fig. \ref{fig:broadcast} and Fig. \ref{fig:latencycalculations}). In case of low PERs, this nearly corresponds to half the ADR latency as the IPD time is very short. 

As it can be seen, a VLC signalling system in a realistic road scenario can yield decisive advantages in terms of total braking distance, which can turn out to be paradigmatic in case of short distances or high speeds, where the car can stop \emph{before} the crossing instead of getting \emph{into the middle} of it (see Table \ref{tab:veh}). 
It is worth to note that this improvement in terms of stopping distance, could be not granted by RF-based communication technologies, for which the IEEE 802.11p standard for ITS envisions maximum average latencies of $100$\,ms, 10 times larger than the ones observed in our worst case scenario tests (0.999 probability,  50 m, 230 kBd, giving latencies below 10 ms, see Fig. \ref{fig:success_prob}). 
Noticeably then, also total relay ADR latencies, reported in Table \ref{tab:veh}, still correspond to reaction distances well below 0.1 m even for speeds as high as 90 km/h. This important point suggests our architecture as a very promising candidate for implementation of low-latency cooperative ITS schemes such as short-distance vehicle platooning, where effective reaction distances well below the 1m range represent a challenging, key target.

\section{Conclusions}\label{sec:conc}
In this work we have constructed and tested a combined VLC Infrastructure-to-Vehicle-to-Vehicle architecture using as a transmitter an LED based regulatory traffic light received on a conventional photodiode driving an ultra-fast active decode-and-relay V2V stage. Our system is based on a low-cost, open-source microcontroller platform (Arduino DUE) and is fully compliant with the IEEE 802.15.7 specification. Our architecture has been tested to a rate of 230 kBd with a Manchester encoding scheme.

We evaluated the performance of our system by a direct measurement of the PER for distances up to 50 m, approaching a value of $10^{-5}$ in optimal conditions at our highest rate of 230 kBd. We performed a statistical analysis taking into account the distribution of errors in
the transmission chain as a function of relevant experimental parameters, to obtain a model predicting statistically-averaged latency
values (SAL) below 1 ms with 99.9\% probability for PER $\lesssim 5\times10^{-3}$. This makes our system already integrable in the new 5G standard protocols.

Our work demonstrates, for the first time, the possibility to attain sub-ms active decode and relay I2V2V communication by integration of VLC technology in real road signaling infrastructures, and the measured performances can redefine new safety standards for a new generation of ITS and cooperative ITS implementations, such as automatic braking, collision avoidance, car platooning, as well as continuous information exchange in VANETs applications \cite{Vegni13}.
For example our VLC signalling system in a realistic road scenario can yield decisive advantages over conventional RF-based technologies in terms of total braking distance, especially in case of short distances or high speeds, where the car can stop \emph{before} the crossing instead of getting \emph{into the middle} of it.

\section*{Acknowledgment}
This work has been performed in collaboration with the Visible Light Communications Research laboratory (VisiCoRe, https://visicore.dinfo.unifi.it/), a partnership of University of Florence, LENS, and CNR-INO.   

Authors would like to thank the company ILES srl (http://www.ilessrl.com) for supplying the traffic-light and for the important support given during the measurements campaign.  

Authors would also acknowledge financial support by Fondazione CR Firenze through project 2016.0792 VisiCom.

\ifCLASSOPTIONcaptionsoff
  \newpage
\fi

\bibliographystyle{IEEEtran}

\bibliography{Bibliography}
%




%



\begin{IEEEbiography}[{\includegraphics[width=1in,height=1.25in,clip,keepaspectratio]{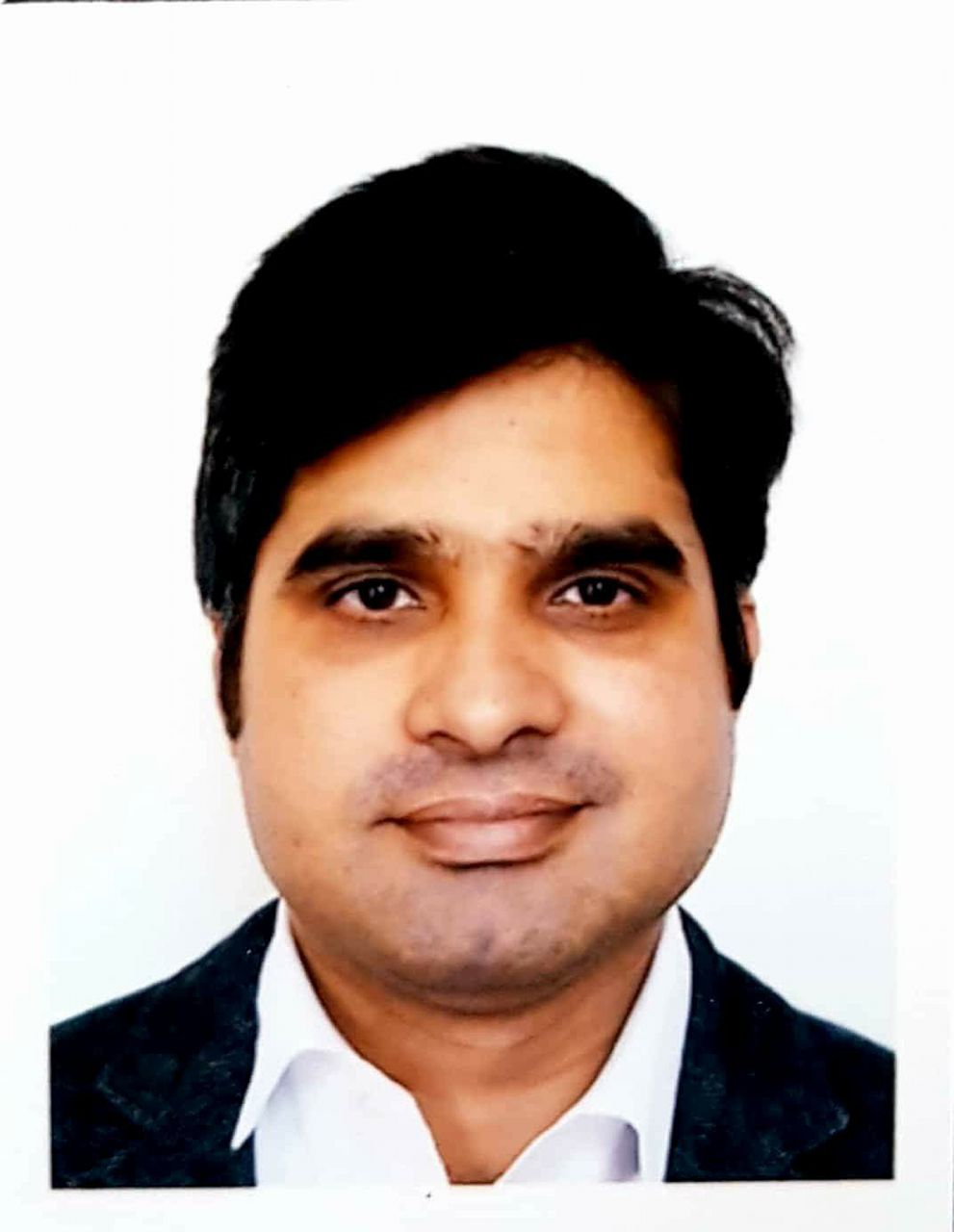}}]{Tassadaq Nawaz} received his M.S. degree in Telecommunications Engineering from  Politecnico di Torino, Italy, in 2013 and Ph.D. in Electronic and Telecommunication Engineering from the University of Genoa, Italy, in 2018. He is currently working as a Postdoctoral researcher at National Research Council-Institute of Optics (CNR-INO), Sesto Fiorentino (FI), Italy. His research interests include communications and signal processing, cognitive radios, visible light communications, physical layer security, and communication electronic warfare solutions.
\end{IEEEbiography}

\begin{IEEEbiography}[{\includegraphics[width=1in,height=1.3in,clip,keepaspectratio]{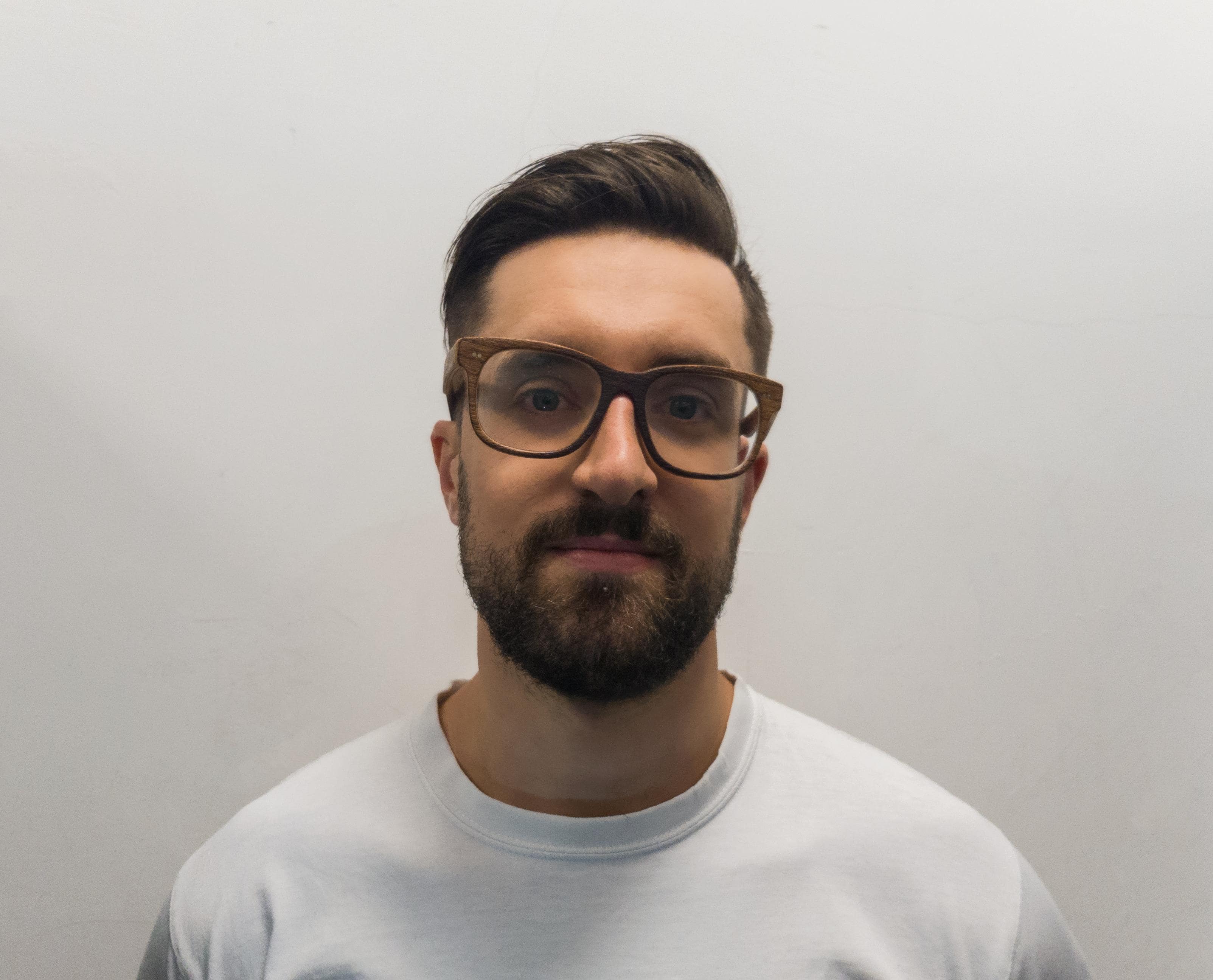}}]
{Marco Seminara} was born in Bagno a Ripoli (Fi), Italy, in 1990. In 2018, he received from the University of Florence the master’s degree in Astrophysics and Physics Sciences working on ultracold quantum gases (Atomic Physics). Right now, he is a PhD Student in: “International Doctorate in Atomic and Molecular Photonics” at European Laboratory for Non-linear Spectroscopy (LENS) – University of Florence. His research is focused on Visible Light Communication in automotive sector (I2V, V2V, I2V2V).
\end{IEEEbiography}

\begin{IEEEbiography}[{\includegraphics[width=1in,height=1.25in,clip,keepaspectratio]{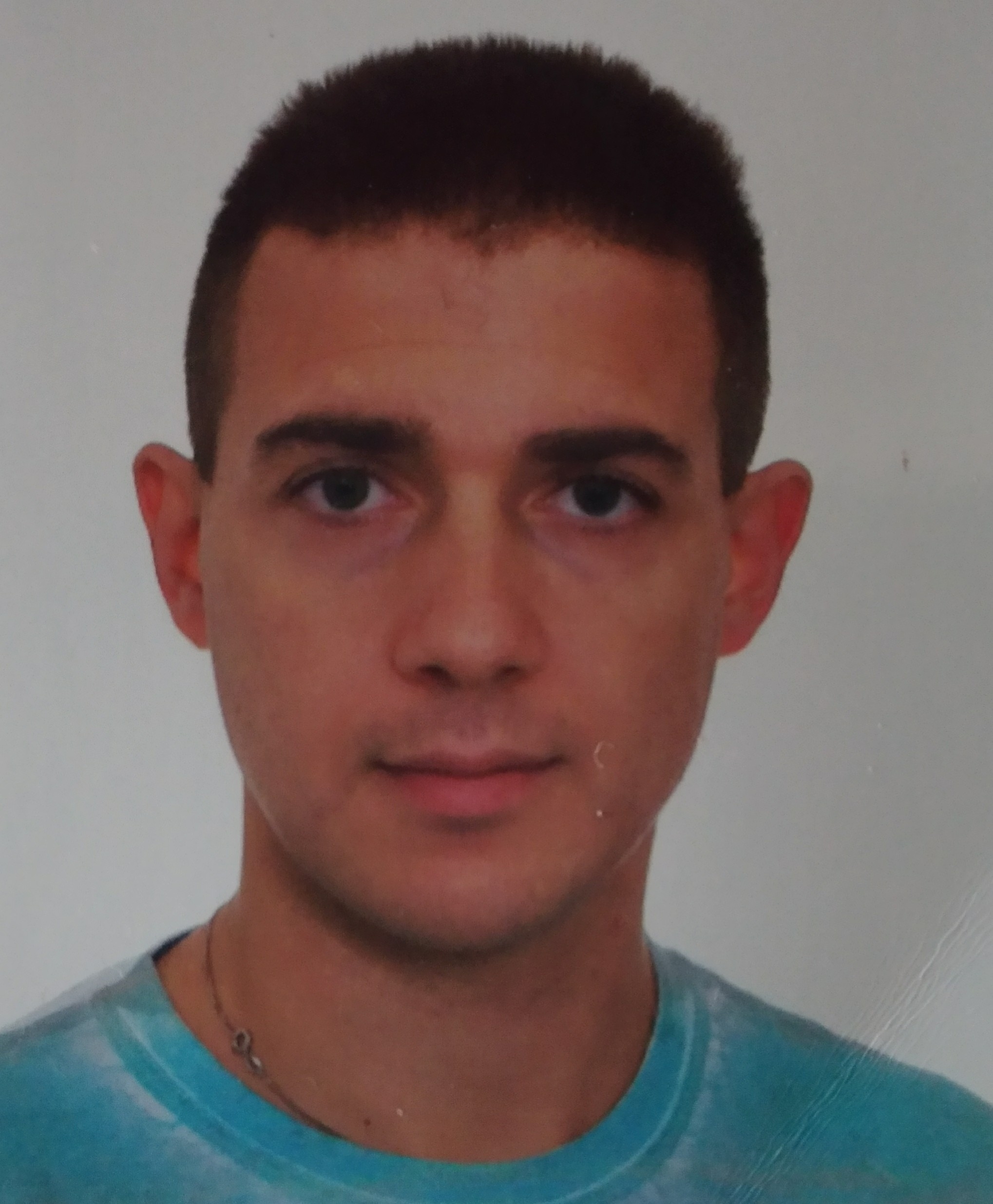}}]{Stefano Caputo} was born in Florence, Italy, in 1984. He received the Dr. Eng. Degree (Laurea) in Mechanical Engineering from the University of Florence (Italy) in 2016 and, currently, he is Ph.D. student in Telecommunications Engineering in the University of Florence (Italy). During the 2010-2011, he spent an 18-month period of work as Mechanical Engineer for design CNC Machine. His main research areas include theoretical modelling, algorithm design and real measurements, mainly focused in the following fields: physical-layer security and light cryptography, sensors and V2V/V2I communication in Automotive field, visible light communications, localization, body area networks, molecular communications.
\end{IEEEbiography}

\begin{IEEEbiography}[{\includegraphics[width=1in,height=1.25in,clip,keepaspectratio]{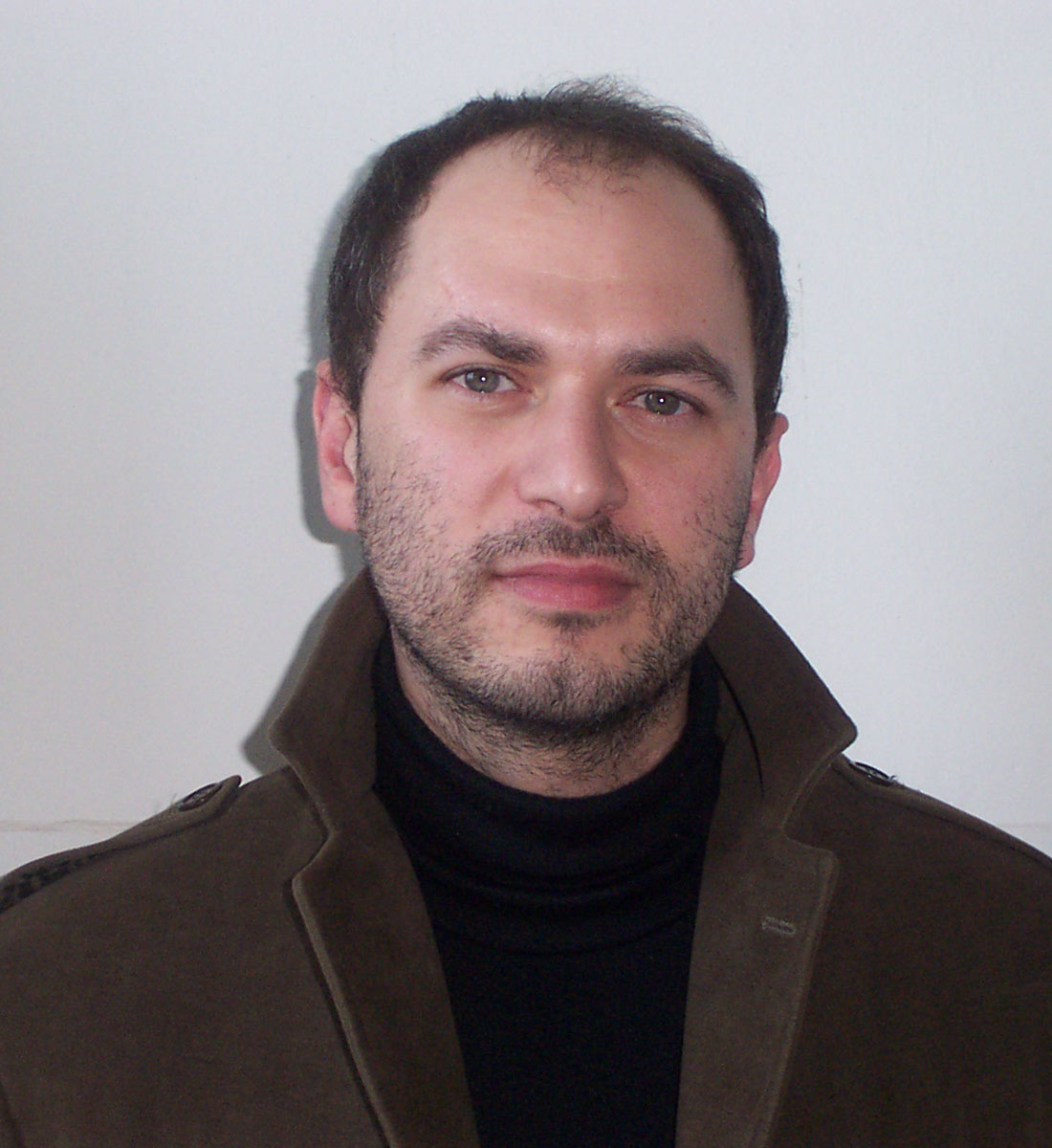}}]{Lorenzo Mucchi} (M'98-SM'12) received the Laurea in telecommunications engineering and the Ph.D. in telecommunications and information society from the University of Florence, Italy, in 1998 and in 2001, respectively. He is an Associate Professor at the University of Florence, Italy. His research interests involve theory and experimentation of wireless systems and networks including physical-layer security, visible light communications, ultra-wideband techniques, body area networks, and interference management. Dr. Mucchi is serving as an associate editor of IEEE Communications Letter and IEEE Access, and he has been Editor-in-Chief for Elsevier Academic Press. He is a member of the European Telecommunications Standard Institute (ETSI) Smart Body Area Network (SmartBAN) group (2013) and team leader of the special task force 511 (2016) 'SmartBAN Performance and Coexistence Verification'. He has been lead organizer and general chair of IEEE and EAI international conferences.  
\end{IEEEbiography}

\begin{IEEEbiography}[{\includegraphics[width=1in,height=1.25in,clip,keepaspectratio]{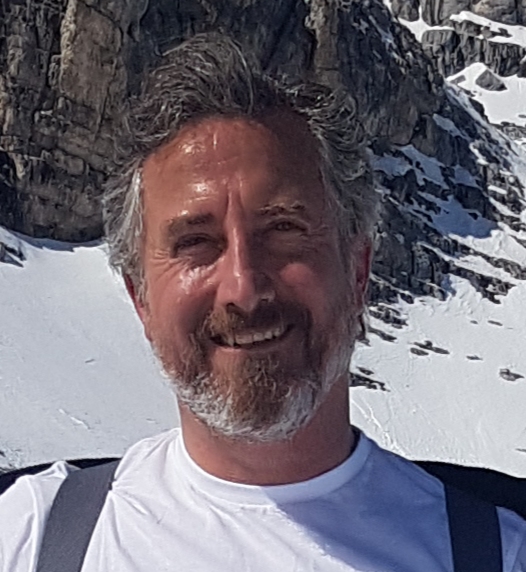}}]{Francesco S. Cataliotti}
Francesco S. Cataliotti is Professor of AMO Physics at University of Firenze. His research interests are centered on quantum technologies with atoms, solid state systems, molecules and photons. He is also concerned with classical sensors and secure communication schemes.  
\end{IEEEbiography}

\begin{IEEEbiography}[{\includegraphics[width=1in,height=1.25in,clip,keepaspectratio]{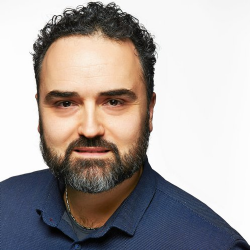}}]{Jacopo Catani}
Jacopo Catani is permanent Researcher at National Institute of Optics - CNR (CNR-INO) in Sesto Fiorentino, and research associate of LENS. He is involved since 2007 in research on quantum gases and quantum simulators using complex optical (Laser and LED) light sources and systems. Since 2016 he is leader of research group in Optical Wireless Communications and Visible-light Communications at CNR-INO. He is head of technology transfer network of CNR-INO since 2015.
\end{IEEEbiography}

\begin{IEEEbiographynophoto}{}
\end{IEEEbiographynophoto}
\begin{IEEEbiographynophoto}{}
\end{IEEEbiographynophoto}
\begin{IEEEbiographynophoto}{}
\end{IEEEbiographynophoto}
\begin{IEEEbiographynophoto}{}
\end{IEEEbiographynophoto}




\end{document}